# Exact pairing correlations in one-dimensional trapped fermions with stochastic mean-field wave-functions

O. Juillet and F. Gulminelli
LPC/ISMRA, Boulevard du Maréchal Juin, F-14050 Caen Cedex, France

Ph. Chomaz
GANIL, BP 5027, F-14076 Caen Cedex 5, France

The canonical thermodynamic properties of a one-dimensional system of interacting spin-1/2 fermions with an attractive zero-range pseudo-potential are investigated within an exact approach. The density operator is evaluated as the statistical average of dyadics formed from a stochastic mean-field propagation of independent Slater determinants. For an harmonically trapped Fermi gas and for fermions confined in a 1D-like torus, we observe the transition to a quasi-BCS state with Cooper-like momentum correlations and an algebraic long-range order. For few trapped fermions in a rotating torus, a dominant superfluid component with quantized circulation can be isolated.

05.30.Fk, 03.75.Ss, 71.10.Pm

Since the pioneering works[1] on the realization of atomic Bose-Einstein condensates (BEC), a great variety of experimental investigations has probed the macroscopic coherence of confined quantum gases in the superfluid regime. For example, by rotating an asymmetric potential, quantized vortex and vortex lattices have been generated in atomic condensates[2]. With the recent progresses on cooling of trapped fermionic atoms[3], one major observation to be detected is now the predicted[4] Bardon-Cooper-Schiffer

(BCS) transition to a paired-fermion superfluid state. The use of Fesbach resonance also opens up the possibility to achieve strong-coupling fermionic superfluidity and the associated BCS-BEC crossover[5]. Due to the experimental realization of very anisotropic trapping geometries[6], the study of the peculiar properties of ultra-cold atomic gases in reduced dimensions is also a topic of growing interest[7]. In this letter, we investigate Cooper-pair formation in a 1D trapped Fermi gas of two hyperfine states $\sigma = +, -$. With a delta-functional two-body interaction, exact eigenfunctions can be obtained via the Bethe ansatz only for the spatially homogeneous system[8]. Here, we perform exact calculations at finite temperature of local and non-local correlation functions in the framework of our stochastic reformulation of the N-fermion problem with binary interactions[9].

An interacting 1D trapped Fermi gas can be modeled by the lattice Hamiltonian

$$H = \delta x \sum_{x \sigma} \Psi_\sigma^+(x) \hat{h}_o \Psi_\sigma(x) - g \delta x \sum_x \Psi_+^+(x) \Psi_-^+(x) \Psi_+(x) \Psi_-(x) \qquad (1)$$

where the coordinates $x$ run on a grid of an even number $N$ of points with periodic boundary conditions. The cell size $\delta x$ is smaller than the macroscopic physical scales. The field operators $\Psi_\sigma(x)$ satisfy the anticommutation relations $\left[\Psi_\sigma^+(x), \Psi_{\sigma'}(x')\right]_+ = \delta_{xx'} \delta_{\sigma\sigma'}/\delta x$ and can be expanded on the plane wave basis according to $\Psi_\sigma(x) = \sum_k a_\sigma(k) e^{ikx}/\sqrt{L}$ where $k = 2\pi n/L$ with the integer $n$ running from $-N/2$ to $N/2-1$, $L = N \delta x$ being the length of the lattice. The one-body Hamiltonian in the confining potential $U(x)$ is $\hat{h}_o = p^2/2m + U(x)$, where $m$ is the atomic mass and $p$ the single-particle momentum operator. Two-body interactions are modeled by a discrete delta pseudo-potential with a coupling constant $g$. The discrete Hamiltonian (1) can also represent interacting fermions trapped in a ring of radius $R$: taking $L = 2\pi$, $x$ then correspond to the azimuthal angle, $p$ to the angular momentum, and $\hat{h}_o = p^2/2mR^2 - \Omega p$ where the contribution $-\Omega p$ is specific to a trap rotating at frequency $\Omega$ and described in the rotating frame. Introducing any single-particle basis $\{|\varphi_\sigma\rangle\}$ for each spin $\sigma$, one can define a basis of the Hilbert space with a fixed number $N_\sigma$ of fermions in each spin state by considering all Slater determinants $|\Phi\rangle = |\Phi_+\rangle |\Phi_-\rangle$ where $|\Phi_\sigma\rangle$ correspond to the occupation of $N_\sigma$ states: $\{\varphi_\sigma^{(i)}, i = 1, \ldots, N_\sigma\}$. Taking advantage of the closure relation $1 = \sum_\Phi |\Phi\rangle \langle \Phi|$, the unnormalized canonical equilibrium density matrix at temperature $k_B T = 1/\beta$ thus reads

$$e^{-\beta H} = \sum_\Phi e^{-\beta H/2} |\Phi\rangle \langle \Phi| e^{-\beta H/2} \qquad (2)$$

To optimize the exploration of the closure relation, in actual simulations, we use the eigenbasis of the single-particle Hamiltonian $\hat{h}_o$ and bias the expansion (2) with the Boltzmann factor associated to $\hat{h}_o$:

$$e^{-\beta H} = \sum_\Phi e^{-\beta E_o(\Phi)} \frac{e^{-\beta H/2} |\Phi\rangle \langle \Phi| e^{-\beta H/2}}{e^{-\beta E_o(\Phi)}} \qquad (3)$$

We have recently shown[9] that the density matrix $e^{-\beta H/2} |\Phi\rangle \langle \Phi| e^{-\beta H/2}$ associated with the imaginary time propagation of each state $|\Phi\rangle$ can be exactly reconstructed by the coherent average of dyadics $|\Phi_1(\beta/2)\rangle \langle \Phi_2(\beta/2)|$. The two independent Slater determinants $|\Phi_\alpha(\beta/2)\rangle (\alpha = 1, 2)$ result from the evolution with a mean-field Hamiltonian supplemented with a one-particle-one-hole Itô noise. Explicitly[9],

$$e^{-\beta H} = \sum_\Phi e^{-\beta E_o(\Phi)} E\left(e^{S(\beta)} |\Phi_1(\beta/2)\rangle \langle \Phi_2(\beta/2)|\right) \qquad (4)$$

with the following stochastic differential equations in imaginary-time $\tau$ for the occupied single-particle orbitals

$$d\varphi_{\sigma\alpha}^{(i)}(x,\tau) = -d\tau[h_0 + g\rho_{-\sigma\alpha}(x,\tau)]\,\varphi_{\sigma\alpha}^{(i)}(x,\tau)$$
$$+ Q_{\sigma\alpha}(\tau) \sum_{k<0} \sqrt{\frac{-g}{L(1+\delta_{k,-\pi/\delta x})}} \left(e^{ikx} dZ_k + hc\right) \varphi_{\sigma\alpha}^{(i)}(x,\tau) \quad (5a)$$

The associated phase evolves according to

$$dS(\tau) = g d\tau \sum_{x\sigma\alpha} \delta x \, \rho_{\sigma\alpha}(x,\tau) \, \rho_{-\sigma\alpha}(x,\tau)/2 \quad (5b)$$

and the initial conditions are

$$\varphi_{\sigma\alpha}^{(i)}(x,0) = \varphi_\sigma^{(i)}(x), \ S(0) = \beta E_o(\Phi) \quad (6)$$

Here, $E(...)$ denotes the ensemble average; $\rho_{\sigma\alpha}(x,\tau)$ is the spatial density associated to the unnormalized Slater determinant $|\Phi_\alpha(\tau)\rangle$. The density-dependent part of the deterministic evolution is the self-consistent Hartree-Fock. $Q_{\sigma\alpha}(\tau)$ is the orthogonal projector to the Fermi sea of spin $\sigma$ associated to $|\Phi_\alpha(\tau)\rangle$. Writing the residual interaction as $\frac{g}{2}\sum_k n(k)n^+(k)$, where $n(k)$ is the discrete Fourier transform of $\sum_\sigma \Psi_\sigma^+(x)\Psi_\sigma(x)$, allows to linearize the two-particles-two-holes excitations by the introduction of complex Wiener processes $Z_k$ obeying to Itô rules :

$$E(dZ_k) = E(\overline{dZ_k}) = 0, dZ_k\, dZ_{k'} = \overline{dZ_k}\,\overline{dZ_{k'}} = 0,\ dZ_k\,\overline{dZ_{k'}} = \delta_{kk'}\,d\tau \quad (7)$$

With the Hamiltonian (1) and with single-particle orbitals that does not mix different spin states, the reformulation (4-6) of the fermionic many-body problem is very close to the formalism recently introduced in the bosonic case[10]. Concerning fermions, the stochastic mean-field scheme can be combined with standard quantum Monte-Carlo algorithms where stochastic paths are generated according to their real weight in the partition function. However, the drawback of the sign-problem would not be solved. Here we proceed in a different way by computing separately the Boltzmann operator, with the representation (4), and its trace.

We first consider a 1D system of $N_+ = N_- = 10$ harmonically trapped fermions interacting via an attractive pseudo-potential. The coupling constant is $g = -2\,\hbar\omega\,a_o$ where $\omega$ is the trap frequency and $a_o = \sqrt{\hbar/m\omega}$. Calculations were performed on a grid of $N = 64$ or $32$ points with the lattice spacing $\delta x = 0.25 a_o$ or $\delta x = 0.4 a_o$ depending on the temperature $k_B T = 2\,\hbar\omega$ or $k_B T = 0.5\,\hbar\omega$. For all the thermal mean values, statistical error bars are less than 5%. The expected development of Cooper pairing at low temperature clearly emerge from the exact stochastic mean-field calculations, as shown in fig.1 in terms of the following momentum correlation function :

$$\chi(k,k') = \langle a_+^+(k)\,a_-^+(k')\,a_-(k')\,a_+(k) \rangle - \langle a_+^+(k)\,a_+(k) \rangle \langle a_-^+(k')\,a_-(k') \rangle \qquad (8)$$

where $\langle\;\rangle$ is the average at equilibrium. As already observed in grand-canonical calculations with small coupling constants[11], only correlations at $k' = k$ or $k' = -k$ around the Fermi surface are dominant (see fig.1). The first ones reflect the formation of co-oscillating pairs of fermions in different spin states. As the temperature decreases, these semi-classical correlations gradually disappear and at $k_B T = 0.5\,\hbar\omega$, the behavior of $\chi(k,k)$ reveals an anticorrelation at equal momentum. On the contrary, correlations between spin-up and spin-down particles, with opposite momenta in the vicinity of the Fermi surface, grow when the temperature is lowered. At $k_B T = 0.5\,\hbar\omega$, these correlations dominate and probe the transition to a Cooper paired state. The quantum coherence properties of this state can now be investigated via the non-local correlation function

$$\gamma(x) = \sum_{x'} \delta x \left( \langle \Psi_+^+(x) \Psi_-^+(x) \Psi_-(x+x') \Psi_+(x+x') \rangle \right.$$
$$\left. - \langle \Psi_+^+(x) \Psi_+(x+x') \rangle \langle \Psi_-^+(x) \Psi_-(x+x') \rangle \right) \quad (9)$$

In the usual BCS picture, the condensation of Cooper pairs induces a long-range order (LRO) in the system, signaled by a nonvanishing correlation function $\gamma(x)$ in the asymptotic regime $x \to \infty$. However, in reduced dimensions, such a scenario is generally perturbed by quantum fluctuations that inhibit any LRO in the strict sense at finite temperature[12]. For 1D harmonically trapped fermions, our exact calculations reveal a similar behavior. The results are presented in fig.2 where $\gamma(x)$ is compared with the correlation function between spin-up and spin-down densities :

$$\kappa(x) = \sum_{x'} \delta x \left( \langle \Psi_+^+(x) \Psi_+(x) \Psi_-^+(x+x') \Psi_-(x+x') \rangle \right.$$
$$\left. - \langle \Psi_+^+(x) \Psi_+(x) \rangle \langle \Psi_-^+(x+x') \Psi_-(x+x') \rangle \right) \quad (10)$$

For $k_B T = 2\, \hbar\omega$, where Cooper pairs coexist with semi-classical correlations, the coherence function $\gamma(x)$ rapidly decays to zero at a distance comparable to the pair correlation length. In the Cooper paired state obtained at temperature $k_B T = 0.5\, \hbar\omega$, $\gamma(x)$ decays as a power law with an exponent $\alpha \approx 1.4$. In contrast with the BCS state which always displays quantum coherence, 1D fermionic paired states in an harmonic trap only exhibit an algebraic LRO.

In a true BCS or BEC state, the long-range order induce a superfluid behavior that can be revealed by generating topological defects like quantized vortices. Moreover, as well known from bi-dimensional systems undergoing a Kosterlitz-Thouless transition, a quasi-LRO can be sufficient to induce superfluidity. Motivated by the famous "rotating bucket experiment with superfluid helium liquid and by recent works on atomic condensates[2], we now investigate a small sample of $N_+ = N_- = 5$ interacting fermions in a ring-like trap rotating at various frequencies $\Omega$. The strength of the δ-pseudopotential interaction is $g = -2\, \hbar\omega$ where $\omega = \hbar/mR^2$ is the frequency associated to the trap. All the calculations have been performed in the rotating frame for a temperature $k_B T = 0.5\, \hbar\omega$ on a $N = 16$ points grid and $k_B T = 2\, \hbar\omega$ with $N = 32$ points. The statistical error bars on the thermal mean values are less than 5%. Without rotation, low-temperature equilibrium states display the same features as those shown in fig.1,2 for harmonically trapped fermions: dominant Cooper pairing correlations around the Fermi surface and algebraic-LRO (see fig.3). When the ring is set into rotation, we analyze the principal pairing mode identified as the eigenvector $\Delta(x)$ of the two-body correlation matrix

$$\rho_2(x',x) = \langle \Psi_+^+(x)\, \Psi_-^+(x)\, \Psi_-(x')\, \Psi_+(x') \rangle - \langle \Psi_+^+(x)\, \Psi_+(x') \rangle \langle \Psi_-^+(x)\, \Psi_-(x') \rangle \quad (11)$$

corresponding to the largest eigenvalue. Such a procedure combines Yang's definition of the order parameter for the superfluid phase transition in macroscopic fermionic systems [13] and the finite-size corrections introduced for ultra-small superconducting grains[14]. A dominant superfluid component should manifest as an irrotational behavior via the Hess-Fairbank effect[15] in the pairing field $\Delta(x)$ when the trap is rotating slowly. Our results are summarized in fig.3 and effectively confirm this behavior: at $\Omega = 0.125,\ 0.25\, \omega$, no rotation of the pairing mode $\Delta(x)$ appears. Indeed, at this low frequency, $\Delta(x) = \sqrt{\rho_\Delta}\, Exp(i\Omega_\Delta x)$ is real in all grid points, which implies a zero angular frequency $\Omega_\Delta$. As expected in a superfluid scenario, when the frequency $\Omega$ of the rotation drive is increased, the mean angular momentum $L_\Delta = \hbar\Omega_\Delta$ associated to $\Delta(x)$ exhibits plateaus of the quantized circulation (fig.3, bottom part). It is interesting to remark that the dominant mode $\Delta(x)$ coexists with at least another non-negligible pairing field rotating at a different frequency (fig.3, bottom part). When the temperature increases, all the pairing eigenmodes of the correlation matrix $\rho_2$ have a comparable weight and the dominant superfluid behavior is destroyed.

In conclusion, we have performed exact stochastic mean-field calculations for 1D trapped fermions at finite temperature. With attractive binary interactions, we have observed the transition to a fermion-paired state characterized by dominant Cooper pairing correlations, an algebraic long-range order and a superfluid component. Even if we are currently limited to a rather small number of particles, these first realistic calculations with stochastic mean-field wave functions illustrate the ability of our method to obtain the equilibrium state of fermionic systems in the canonical ensemble. In addition, the imaginary-time fluctuating mean-field propagation used here can be coupled to a real-time one to get the exact dynamics of a initial finite temperature state. Finally, the same approach can be used with trial wave-functions different from Slater determinants: for example, investigations on macroscopic fermionic systems dominated by pairing correlations are under development with stochastic BCS wave-functions.

We acknowledge fruitful discussions with J. Dalibard, Y. Castin and I. Carusotto. We also gratefully thank R. Fresard and P. van Isacker for a careful reading of the manuscript.

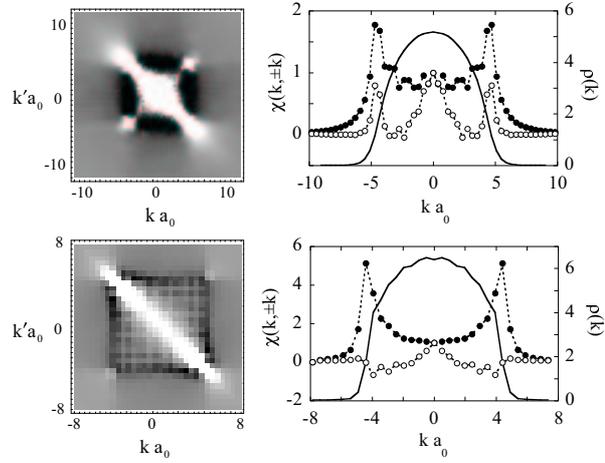

**FIG.1. Cooper pairing correlations in a 1D system of $N_+ = N_- = 10$ interacting fermions in an harmonic trap at different temperatures $k_B T = 2\ \hbar\omega$ (top) and $k_B T = 0.5\ \hbar\omega$ (bottom). Left part: density plot of the momentum correlation function $\chi(k,k')$. Right part: momentum density profiles $\rho(k)$ (solid line) and correlations $\chi(k,\pm k)$ (dashed lines) relative to their value for $k=0$ (open circles: $\chi(k,k)$, black disks: $\chi(k,-k)$).**

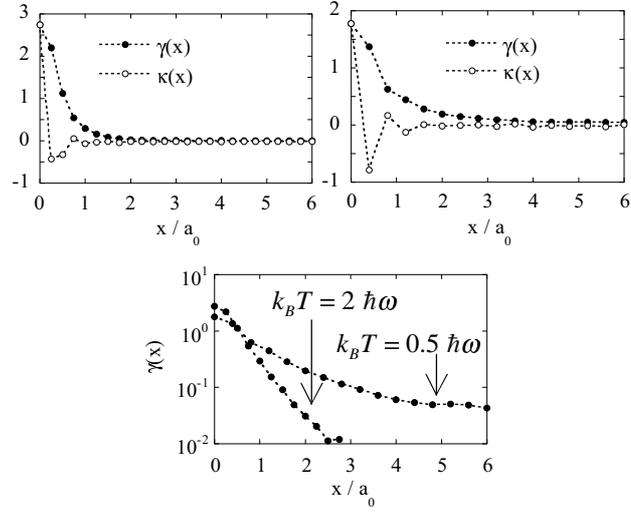

**FIG.2.** Upper part : exact non-local correlation functions $\gamma(x)$ and $\kappa(x)$ for $N_+ = N_- = 10$ **harmonically trapped fermions at different temperatures** $k_B T = 2\,\hbar\omega$ **(left) and** $k_B T = 0.5\,\hbar\omega$ **(right). Lower part: comparison of the coherence function** $\gamma(x)$ **for the two considered temperatures.**

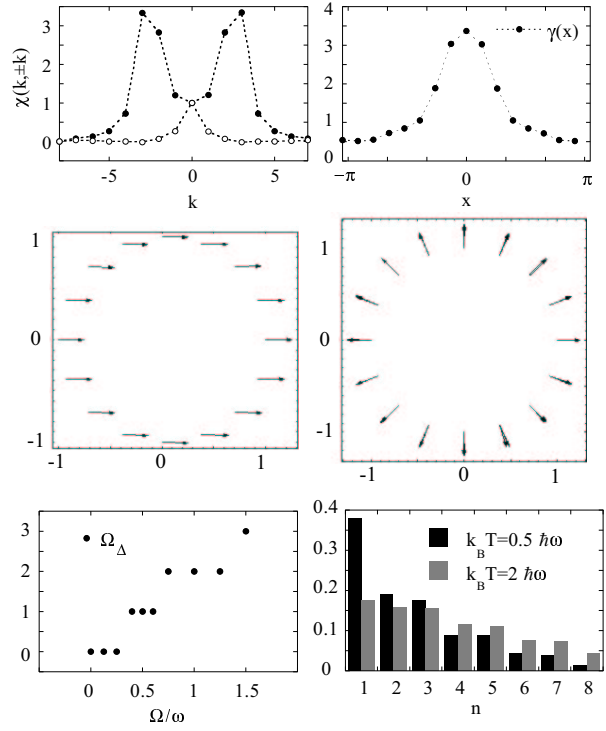

**FIG.3. Stochastic Hartree-Fock calculations of the thermal equilibrium state of $N_+ = N_- = 5$ interacting fermions confined in a rotating ring of radius $R$ at a temperature $k_B T = 0.5\, \hbar\omega$ with $\omega = \hbar/mR^2$.**

**Upper part : coherence function $\gamma(x)$ and momentum correlations $\chi(k, \pm k)$ relative to their value for $k = 0$ (open circles: $\chi(k,k)$, black disks: $\chi(k,-k)$) for the non-rotating trap. Middle part: representation of the dominant pairing mode $\Delta(x)$ in the complex plane for each point $x$ considered as a polar angle on the ring. This vector field representation is shown for the frequency of the rotating drive $\Omega = 0.125, 0.25\, \omega$ (left) and $\Omega = 0.4, 0.5, 0.6\, \omega$ (right). The different fields are indistinguishable at the scale of the figure. Lower part: angular frequency of rotation $\Omega_\Delta$ associated to the velocity field of the principal pairing mode (left);**

distribution of the eigenvalues $\lambda_n$ (relative to their sum) of the correlation matrix $\rho_2$ at different temperatures $k_B T = 0.5,\ 2\ \hbar\omega$ and for a frequency of rotation $\Omega = 0.5\ \omega$ **(right).**